\documentclass[twocolumn,showpacs,preprintnumbers,prb,amsmath,amssymb,superscriptaddress]{revtex4}

\usepackage{graphicx}
\usepackage{dcolumn}
\usepackage{bm}

\begin{document}


\title{Empirical potential study of phonon transport in graphitic ribbon}
\author{Takahiro Yamamoto}\email[e-mail: ]{takahiro@rs.kagu.tus.ac.jp}
\affiliation{
Department of Physics, Faculty of Science, Tokyo University of Science,
1-3 Kagurazaka, Shinjuku-ku, Tokyo 162-8601, Japan}
\affiliation{
CREST, Japan Science and Technology Agency, 4-1-8 Honcho Kawaguchi, Saitama, Japan
}
  
\author{Kazuaki Mii}
\affiliation{
Department of Physics, Faculty of Science, Tokyo University of Science,
1-3 Kagurazaka, Shinjuku-ku, Tokyo 162-8601, Japan}
\author{Kazuyuki Watanabe}
\affiliation{
Department of Physics, Faculty of Science, Tokyo University of Science,
1-3 Kagurazaka, Shinjuku-ku, Tokyo 162-8601, Japan}
\affiliation{
CREST, Japan Science and Technology Agency, 4-1-8 Honcho Kawaguchi, Saitama, Japan }

\date{\today}

\begin{abstract} 
The thermal properties of graphitic ribbon are investigated based on Brenner's empirical potential. The reliability and usefulness of the empirical potential to address the thermal properties of covalent-bonded carbon nanostructures are verified through a comparison of phonon dispersion relations and the density of states of carbon nanotubes with first-principles calculations. The analysis reveals unique edge-phonon states that are highly localized at edge carbon atoms of both armchair and zigzag ribbons. Applying the phonon dispersion relations to the Landauer formula of phonon transport, the quantization and universal features of the low-temperature thermal conductance of graphitic ribbon are elucidated, and it is found that the width of the quantization plateau in the low-temperature region is inversely proportional to the ribbon width.
\end{abstract}

 
\maketitle

\section{Introduction}
Carbon is unique among all the elements in that it can form strong covalent bonds of $sp$, $sp^2$, and $sp^3$ hybridizations, depending on the atomic geometry. Owing to this strong covalent bond behavior, carbon allotropes exhibit remarkable thermodynamic and thermal transport properties. For example, crystalline diamond has the lowest specific heat and highest thermal conductivity of all solids. In contrast to bulk crystals, however, it is difficult to measure thermal quantities of carbon nanostructures such as fullerenes~\cite{rf:kroto} and carbon nanotubes,~\cite{rf:iijima} as it is necessary for measurement to ensure complete thermal isolation from surrounding materials. The thermal transport properties of carbon nanostructures are the key to controlling the performance and stability of future nanodevices, and it is now widely accepted that thermal management of nanostructures cannot be avoided as the size of devices reaches the nanometer scale. Therefore, it has become particularly important to be able to understand and predict the thermal transport properties of carbon nanostructures through theoretical and computational analyses.

The thermal transport properties of carbon nanostructures have been widely investigated by molecular dynamics (MD) methods based on the empirical potential developed by Brenner,~\cite{rf:brenner} which is particularly suitable for potential functions of carbon materials. Recent MD studies on thermal transport in single-wall carbon nanotubes (SWNTs) have predicted that the SWNTs at room temperature have relatively high thermal conductance, comparable to that of crystalline diamond.~\cite{rf:Berber-K-T,rf:Che-C-G,rf:Osman-S,rf:Maruyama,rf:Zhang-S-W-G} High thermal conductance has also been observed experimentally using micrometer-sized ropes of SWNTs.~\cite{rf:hone} However, the investigation of thermal transport phenomena at low temperatures is beyond the capability of classical MD methods, as the low-temperature thermal conductance is predicted to exhibit quantum effects. Therefore, it is desirable to develop a reliable scheme that is appropriate for the description of low-temperature thermal transport in carbon nanostructures. 

The thermal transport properties depend critically on the phonon dispersion relations of systems. The phonon dispersion relations are obtained by diagonalizing the dynamical  matrix containing force constants between any two atoms in a system of interest. The force-constant matrix is derived from the second derivative of a potential function with respect to atom coordinates. This work employs the Brenner potential, which has been verified to be adequate for the structural study of carbon materials including diamond surfaces.~\cite{rf:brenner} The dynamical  matrix is derived from the second derivative of the Brenner potential, leading to the phonon dispersion relations of the carbon materials. The phonon dispersion relations for SWNTs are calculated first in this paper to verify the transferability of the Brenner potential and to demonstrate the usefulness of the model. Compared to {\it ab initio} force-constant methods based on density functional theory, the empirical model is faster in computation and can deal with much larger systems. Most importantly, the phonon dispersion relations obtained are in good agreement with those of {\it ab initio} force-constant methods.~\cite{rf:dubay,rf:ye} 

The empirical force-constant model is applied in this paper to calculation of the phonon transport properties of graphitic ribbon as a potential wiring material in nanoscale electronic devices as well as carbon nanotube applications.~\cite{rf:honma} Recently, various forms of graphitic ribbon have been synthesized toward the realization of nanodevices.~\cite{rf:tera,rf:li,rf:cancado} Polyperinaphthalene is a graphitic ribbon with an armchair-shaped edge (armchair ribbon), including 10 carbon atoms in the unit cell.~\cite{rf:yasu,rf:mura} Owing to edge effects (finite size), graphitic ribbon exhibits remarkable physical properties that are not expected to appear for carbon nanotubes with closed structures. Graphitic ribbon with zigzag-shaped edges (zigzag ribbon) exhibits unique electronic states with energies close to the Fermi level. These states are highly localized at the edge carbon atoms, and are defined as edge states, but the states do not appear in the armchair ribbons.~\cite{rf:fujita,rf:nakada,rf:waka-1,rf:miyamoto,rf:ramprasad} Electronic,~\cite{rf:waka-2} magnetic,~\cite{rf:fujita,rf:waka-1,rf:hikihara} optical,~\cite{rf:lin} and field emission~\cite{rf:tada,rf:ara} properties have also been studied extensively with respect to potential edge effects. However, to the best of the authors' knowledge, the phonon transport properties of graphitic ribbon at low temperature have not been studied. Despite a longstanding theoretical interest going back to Peierls' early work,~\cite{rf:peierls2} little progress has been made toward elucidating phonon transport in mesoscopic or nanoscale systems with dimensions much smaller than the mean free path of phonons. Recent advances in nanotechnology have made it possible to measure the thermal conductance by phonons in such small systems and to observe the universal quantum of thermal conductance. The value obtained, $\pi^2 k_B^2 T/3h=9.4\times 10^{-13}$~[W/K],~\cite{rf:schw,rf:yung} is consistent with the value predicted theoretically by Rego and Kirczenow~\cite{rf:rego1} and by other theoretical studies.~\cite{rf:angel,rf:blenc} Very recently, we reported that the quantization of thermal conductance in SWNTs is independent of tube radius or chirality.~\cite{rf:yama1} Motivated by these discoveries, the objective of the present study is to calculate the phonon dispersion relations based on the empirical force-constant model and to demonstrate the universal features of quantized thermal conductance in graphitic ribbon at low temperature.

This paper is organized as follows. In section~\ref{sec:2}, the force-constant method based on Brenner's empirical potential is introduced for investigation of the thermal properties of carbon nanostructures, focusing on SWNTs and graphitic ribbons as primary examples. To verify the transferability of the method, the phonon dispersion relations and the density of states of SWNTs are calculated. In section~\ref{sec:3}, the atomic structures of graphitic ribbon are optimized using the Brenner potential and the phonon dispersion relations of graphitic ribbon are obtained by the empirical force-constant method. In section~\ref{sec:4}, phonon-derived thermal conductance in graphitic ribbon is discussed based on theoretical analysis using the Landauer theory of phonon transport, and the quantization and universal behavior of thermal conductance in graphitic ribbon are presented. The paper is finally summarized in section~\ref{sec:5}. 

\section{Empirical Force-Constant Model\label{sec:2}}
The phonon dispersion relations of a system are obtained by diagonalizing the dynamical  matrix given by
\begin{eqnarray}
D^{j\beta}_{i\alpha}({\mbox{\boldmath $k$}})=\sum_{m}\frac{1}{\sqrt{M_iM_j}}
K^{mj\beta}_{ni\alpha}{\rm exp}\left({\rm i}{\mbox{\boldmath $k$}}\cdot
\left({\mbox{\boldmath $R$}}_m-{\mbox{\boldmath $R$}}_n\right)\right).
\label{eq:d-mat}
\end{eqnarray}
Here, the summation index $m$ runs over all unit cells, ${\mbox{\boldmath $k$}}$ is the wave vector, ${\mbox{\boldmath $R$}}_m$ is the spatial coordinate of the $m$th unit cell, and $M_i$ is the mass of the $i$th atom. $D^{j\beta}_{i\alpha}({\mbox{\boldmath $k$}})$ is independent of $n$ owing to the translational symmetry of the system. The force-constant $K^{mj\beta}_{ni\alpha}$ is defined as the second derivative of the potential with respect to atomic coordinates as follows. 
\begin{eqnarray}
K^{mj\beta}_{ni\alpha}=\frac{\partial^2V}{\partial{r}_{n i\alpha}\partial{r}_{m j\beta}},
\label{eq:K}
\end{eqnarray}
where $V$ is the potential energy of the system, $r_{ni\alpha}$ is the $\alpha$ component $(\alpha=x,y,z)$ of the position vector of the $i$th atom in a unit cell $n$.

For carbon nanostructures, Brenner's empirical potential is adopted for $V$ in Eq.~(\ref{eq:K}). The Brenner potential has been successfully used for MD simulations of carbon nanostructures. Details of this potential can be found in Brenner's original paper.~\cite{rf:brenner} The force constants can of course be obtained for small systems from the Hellmann-Feynmann force computed by first-principles calculations based on density functional theory. However, for large systems, the phonon transport properties must be calculated for systems with a very large phonon mean-free-path, for example, the phonon mean-free-path of SWNTs is of the order of $0.1-1.0~\mu$m.~\cite{rf:hone,rf:javey} The empirical method, on the other hand, can be computed very quickly, making it very useful for computing atomistic quantities in larger systems.

Substituting the Brenner potential into Eq.~(\ref{eq:K}), an analytical form of the force-constant can be obtained in a straightforward manner. The derivation is omitted here for the sake of brevity. The reliability and usefulness of the analytical equation for analysis of the thermal properties of carbon nanostructures is evaluated by comparing phonon dispersion relations for SWNTs determined by this model with the results present in previous studies. As the force constants are only determined for the equilibrium positions of carbon atoms in the system, it is necessary to optimize the atomic geometry of the SWNTs using the Brenner potential. For an SWNT with a chiral vector of ${\mbox{\boldmath $C$}}_h=(10,10)$, where the chiral vector ${\mbox{\boldmath $C$}}_h=(n,m)$ uniquely determines the geometrical structure of the SWNT,~\cite{rf:sai1,rf:dres} the optimal bond length between two carbon atoms is obtained as $1.45$~{\r{A}} using the parameter set given in Table III in Brenner's original paper~\cite{rf:brenner}. This is slightly longer than the typical experimental value of $1.44$~{\r{A}}. Figure~\ref{fig:1}(a) shows the phonon dispersion relations for the $(10,10)$ SWNT, which has 40 carbon atoms in the unit cell and 120 vibrational degrees of freedom. 

Due to mode degeneracies, only 66 distinct phonon branches including 12 nondegenerate branches and 54 doubly degenerate branches appear in Fig.~\ref{fig:1}(a). This result is consistent with the number of distinct phonon branches obtained by point group theory for atoms in the unit cell. The corresponding density of states is shown in Fig.~\ref{fig:1}(b). The small peaks represent van Hove singularities unique to one-dimensional systems.

Four acoustic modes with linear dispersion appear in the low-frequency region of Fig.~\ref{fig:1}(a): a longitudinal acoustic (LA) mode, a twisting acoustic (TW) mode, doubly degenerate transverse acoustic (TA) modes. The sound velocities of the LA, TW, and TA modes for $(10,10)$-SWNT are estimated to be 22.2, 13.3, and 10.3~km/s, respectively. The energy gap of the doubly degenerated lowest optical modes depends only on the tube radius $R$, and decreases approximately according to $\sim 1/R^2$. These results are in good agreement with previously reported phonon dispersion relations for SWNTs determined by a dynamical  matrix method with force-constant parameters for graphene~\cite{rf:sai1} and first-principles calculations.~\cite{rf:dubay,rf:ye}
Therefore, the phonon dispersion relations for other carbon nanostrucures are considered to be reliably obtained using the force-constant method based on Brenner's empirical potential.

\section{Phonon dispersion relations for graphitic ribbon\label{sec:3}}
The empirical force-constant method is employed to calculate the phonon dispersion relations of the two types of graphitic ribbon without  hydrogen termination. The dehydrogenated armchair- and zigzag-ribbons are stable even at high temperatures above 2000K.~\cite{rf:kawai}
Figure~\ref{fig:2} shows the schematic structures of the dehydrogenated armchair- and zigzag-ribbons. Here, $N_a$ and $N_z$ denote the number of dimer lines and zigzag lines, respectively. Before calculating the phonon dispersion relations, the atomic structures need to be fully optimized using the Brenner potential. For the armchair ribbon with $N_a=9$ ($18$ atoms per unit cell), the optimized bond length between two edge atoms is $1.39$~{\r{A}}, which is shorter than the typical bond length of $1.45$~{\r{A}} inside the armchair ribbon. This shortening of the edge bond length appeared for all armchair ribbons calculated in this study, and results from the fact that two atoms at the edges of the armchair ribbon form a triple bond, C$\equiv$C. This triple-bond feature has been already revealed in the first-principles calculations of the electron density of armchair ribbons without hydrogen termination.~\cite{rf:kawai}

The phonon dispersion relations represent key information for identifying the triple bond at the edge of the armchair ribbons experimentally. As a result of the strong triple bond of C$\equiv$C, doubly degenerated branches with frequency of approximately 1800~cm$^{-1}$ at the $\Gamma$ point appear in the phonon dispersion relations. These two in-plane modes correspond to antiphase and in-phase C$\equiv$C stretching motions on both edges of the ribbon. The frequency of $\sim$1800~cm$^{-1}$ is higher than the typical SWNT peak at $\sim$1590~cm$^{-1}$ in the Raman spectrum originating from the Raman-active $E_{2g}$ mode of a graphitic sheet.~\cite{rf:rao} As this is indicative of a Raman shift at this frequency, it is possible to identify the C$\equiv$C bond at the edge of an armchair ribbon based on the Raman spectrum. 

The edge-localized phonon modes appearing as out-of-plane modes of the armchair ribbon are shown in Fig.~\ref{fig:3} for three different ribbon widths, together with the dispersion relations for a graphene sheet projected onto the armchair axis (parallel to the translation vector along the ribbon axis ${\mbox{\boldmath $a$}}$). The upper and lower shaded areas correspond to the optical and acoustic modes of the graphene sheet, respectively. For all the armchair ribbons calculated, four special branches were found outside the region of the projected dispersion relations of a graphene sheet, as indicated by arrows in the figure. The existence of these branches has been already predicted by Igami {\it et al.},~\cite{rf:igami1,rf:igami2} but the detailed properties have yet to be reported. From an examination of the atomic displacements, the phonon states in these four branches appear to correspond to a strongly localized state on the armchair edges. Figure~\ref{fig:4} shows snapshots of the atomic displacements at $k|{\mbox{\boldmath $a$}}|/\pi=1$ for these edge-localized phonon states of the armchair ribbon with $N_a=9$. As shown in Fig.~\ref{fig:3}, two branches occur below the lower shaded area, representing an acoustic modes and a lowest optical mode. In the acoustic mode, both edge-dimers in the unit cell undergo in-phase vibration (Fig.~\ref{fig:4}(a)), while in the lowest optical mode, both edge-dimers exhibit anti-phase vibration (Fig.~\ref{fig:4}(b)). Two other edge-localized modes appear in the region between the upper and lower shaded areas in Fig.~\ref{fig:3}, representing a mode due to two carbon atoms in the anti-phase vibrating edge-dimer and the in-phase vibration of the two edge-dimers on each side in the unit cell, as shown in Fig.~\ref{fig:4}(c), and an edge-localized mode in which both edge-dimers vibrate in anti-phase, as shown in Fig.~\ref{fig:4}(d). 

Figure~\ref{fig:5} shows the phonon dispersion relations of the out-of-plane modes for three zigzag ribbons, together with the dispersion relations (shaded areas) of a graphene sheet projected onto the zigzag axis (parallel to the vector ${\mbox{\boldmath $z$}}$). The upper and lower shaded areas contact at $k|{\mbox{\boldmath $z$}}|/\pi=2/3$. The two branches appearing immediately below the lower shaded area cannot be predicted from the projected dispersion relations of a graphene sheet. The phonon states in these two branches correspond to edge-localized states from observations of the atomic motions in these branches (Fig.~\ref{fig:6}). In the acoustic mode, two edge-atoms in the unit cell exhibit in-phase vibration (Fig.~\ref{fig:6}(a)), while in the lowest optical mode, the two edge-atoms vibrate in anti-phase (Fig.~\ref{fig:6}(b)). 

The lowest-lying phonon dispersion relations governing the low-temperature thermal-transport properties of graphitic ribbon are represented by three acoustic modes in the long-wavelength limit; an LA mode, an in-plane TA mode, and an out-of-plane TA mode. In contrast to SWNTs, the two TA modes are not degenerate because there is no rotational symmetry around the LA-wave direction. Using the empirical force-constant method, the energy gap of the lowest optical mode at the $\Gamma$ point, $\Delta{E}_{\rm op}\equiv\hbar\omega_{\rm op}$, was shown to depend only on the ribbon width $W$ according to the relation $\Delta{E}_{\rm op}\propto{1/W}$. That is, the energy gap is independent of the edge shape. The $\Delta{E}_{\rm op}$ of SWNTs has an approximately quadric radius dependence given by $1/R^2$ for tube radius $R$,~\cite{rf:sai1} in contrast to the $\Delta{E}_{\rm op}\propto 1/R$ relation predicted by the zone-folding approach.~\cite{rf:bene} This difference arises because the zone-folding approach neglects the effects of cylinder geometry or curvature of the SWNT. $\Delta{E}_{\rm op}$ represents an important physical parameter for characterizing the quantization plateau of thermal conductance, as shown in the next section.

\section{Phonon transport in graphitic ribbon\label{sec:4}}
From an experimental fact that the phonon mean-free-path of SWNT is 0.5-1.5$\mu$m in the wide temperature range from 350K to 8K,~\cite{rf:hone}
we can expect the graphitic ribbon is a ballistic phonon conductor with a large phonon mean-free-path of the order of 1$\mu$m at low temperatures.
Consider a situation in which phonons travel ballistically through a graphitic ribbon, connected smoothly between two thermal reservoirs with a small temperature difference. 

It has been shown previously in a study of the thermal conductance of SWNTs~\cite{rf:yama1} that the thermal conductance in such situation is well described by 
\begin{eqnarray}
\kappa(T)=\frac{2k_B^2T}{h}\sum_{m}\left[
G(x_m^{\rm min})-G(x_m^{\rm max})
\right],
\label{eq:kappa}
\end{eqnarray}
based on the Landauer theory of phonon transport~\cite{rf:rego1}. Here, $G(x)$ is defined as 
\begin{eqnarray}
G(x)=\phi(2,e^{-x})+x\phi(1,e^{-x})+\frac{x^2}{2}f(x),
\end{eqnarray}
where $x_m^{\rm min/max}$ is the minimum/maximum value of $x_m\equiv\hbar\omega_m/k_BT$, $\phi(z,s)=\sum_{n=1}^{\infty}(s^n/n^z)$ is the Appel function, and $f(x)=1/(e^x-1)$ is the Bose-Einstein distribution function. Details of the derivation of Eq.~(\ref{eq:kappa}) can be found in the previous report.~\cite{rf:yama1}

Substituting $\omega_m^{\rm min}$ and $\omega_m^{\rm max}$ for the graphitic ribbon into Eq.~(\ref{eq:kappa}) gives the thermal conductance of these two types of graphitic ribbon. Figure~\ref{fig:7}(a) shows the temperature dependence of the thermal conductance of graphitic ribbon with different widths and edge shapes, normalized against a universal value of $3\kappa_0$ ($\kappa_0\equiv \pi^2k_B^2T/3h$ is the universal quantum of thermal conductance). All calculated curves approach unity in the low-temperature limit, indicating that the thermal conductance in graphitic ribbon is quantized at the universal value of $3\kappa_0$, independent of the ribbon width or edge shape. It can be seen from Eq.~(\ref{eq:kappa}) that each acoustic mode with $x_m^{\rm min}=0$ (LA mode and in-plane and out-of-plane TA modes) contributes a universal quantum of $\kappa_0$ to the thermal conductance. 

The quantization plateau seen in Fig.~\ref{fig:7}(a) is broken with increasing temperature, with the thermal conductance beginning to rise as the optical phonons become excited. Since the characteristic energy scale of the graphitic ribbon in this temperature region is the energy gap of the lowest optical mode, $\Delta{E}_{\rm op}$, it would be convenient to introduce a dimensionless temperature $\tau\equiv k_BT/\Delta{E}_{\rm op}$. Taking account of the three acoustic modes and the lowest optical mode and substituting the value of $\omega_m^{\rm min}$ for these branches into Eq.~(\ref{eq:kappa}), the thermal conductance can be given in the following simple form.
\begin{eqnarray}
\frac{\kappa(\tau)}{3\kappa_0}=1+\frac{2}{\pi^2}e^{-1/\tau}
\left(1+\frac{1}{\tau}+\frac{1}{2\tau^2}\right).
\label{eq:scale}
\end{eqnarray}
This equation gives the thermal conductance as being dependent only on $\tau$. As shown in Fig.~\ref{fig:7}(b), all curves in Fig.~\ref{fig:7}(a) collapse onto a single curve of Eq.~(\ref{eq:scale}) once plotted against $\tau$. The single curve rises at $\tau\approx{0.1}$ from the quantization plateau. The width of the quantization plateau with respect to temperature, $\Delta{T}\approx{0.1}\times\Delta{E_{\rm op}}/k_B$, is found to be inversely proportional to the ribbon width $W$, that is, $\Delta{T}\propto{1/W}$. This means that the phonon transport properties of graphitic ribbon at low temperature are independent of the atomic structure such as edge shape. This follows from that fact that the wavelength of phonons as heat carriers is sufficiently high in comparison with the bond length of the graphitic ribbon.

Inspired by recent experiments on phonon transport in SWNTs,~\cite{rf:hone} we previously showed that the phonon-derived thermal conductance of SWNTs at low temperatures is quantized at $4\kappa_0$, corresponding to 4 acoustic modes; the LA mode, two TA modes, and the TW mode.~\cite{rf:yama1} Since there is no TW mode in the phonon dispersion curves of graphitic ribbon, the thermal conductance in graphitic ribbon is smaller than that in SWNTs. In graphitic ribbon, one of the three quanta is carried by edge-localized acoustic phonons. Furthermore, while the quantization plateau width for graphitic ribbon extends in proportion to $1/W$ with decreasing $W$, independent of edge structures, that for SWNTs is proportional to $1/R^2$. Therefore, the width of the quantization plateau for narrow graphitic ribbon is smaller than that for SWNTs with the same number of atoms per unit cell. 

\section{Summary\label{sec:5}}
As a first step toward systematic study of the thermal properties of carbon nanostructures, the phonon dispersion relations and density of states of carbon nanostructures were calculated using a force-constant method based on Brenner's empirical potential, focusing on SWNTs as a primary example. The results obtained are in good agreement with previous results for the phonon dispersion relations of SWNTs determined by first-principles calculations.~\cite{rf:dubay,rf:ye} The empirical force-constant method is therefore considered to be a reliable method for calculations of the phonon dispersion relations for various carbon nanostructures. Using this method, phonon dispersion relations were calculated for graphitic ribbon with armchair- or zigzag-shaped edges, and localized phonon modes were identified at the edges of the ribbons. For armchair ribbons, two in-plane phonon modes with high frequency (near 1800~cm$^{-1}$) are assigned to C$\equiv$C stretching motion at the edges. The out-of-plane TA and lowest optical modes were also shown to be edge-localized phonon states in the armchair and zigzag ribbons. These low-lying modes were found to contribute to the thermal transport properties of graphitic ribbon at low temperature. 

Using the Landauer formula for phonon transport, perfect transmission of an acoustic mode in graphitic ribbon was found to contribute a universal quantum of $\kappa_0=\pi^2k_B^2T/3h$ to thermal conductance. The width of the temperature-dependent quantization plateau was also observed to be inversely proportional to ribbon width. Quantization phenomena of thermal conductance are therefore expected to be experimentally observable in various one-dimensional carbon nanostructures such as carbynes and polyacetylenes as well as SWNTs and graphitic ribbons. The successful application of an empirical force-constant method combined with the Landauer formula in this study revealed the quantized and universal features of phonon-derived thermal conductance in graphitic ribbon, demonstrating the essential role of ballistic edge-phonon modes in these materials.

\begin{acknowledgments}
T.Y. thanks Takazumi Kawai for helpful discussions regarding edge-localized phonons in graphitic ribbon. The present study was supported in part by a Grant-in-Aid (No.~15607018) from the Ministry of Education, Sports, Culture, Science and Technology of Japan.
\end{acknowledgments}

\newpage
\begin{figure}[t]
  \caption{(a) Phonon dispersion curves and (b) density of states for a single-walled
  carbon nanotube with chiral vector $(10,10)$. $|{\mbox{\boldmath $T$}}|$ denotes the 
  magnitude of the translation vector along the tube axis.}
  \label{fig:1}
  \caption{Atomic structure of (a) armchair ribbon ($N_a=7$) and (b) zigzag ribbon 
  ($N_z=4$). Broken rectangle represents a unit cell, and ${\mbox{\boldmath $a$}}$ and 
  ${\mbox{\boldmath $z$}}$ denote the translation vectors along the ribbon axis.}
  \label{fig:2}
  \caption{Phonon dispersion relations of armchair ribbon with $N_a=9$, $11$, and $15$. 
  Shaded areas represent the phonon dispersion relations of a two-dimensional graphitic 
  sheet projected onto the armchair axis. Edge-localized phonon modes are indicated by 
  arrows.}
  \label{fig:3}
  \caption{Atomic displacement of edge-localized phonon states at 
  $k|{\mbox{\boldmath $a$}}|/\pi=1$ in the out-of-plane modes of an armchair ribbon of 
  $N_a=9$ for (a) the transverse acoustic mode, and (b) the lowest-lying optical mode. 
  (c), (d) Atomic displacement of states in the two branches between the upper and lower 
  shaded areas in Fig.~\ref{fig:3}. Solid circles represent edge carbon atoms.}
  \label{fig:4}
  \caption{Phonon dispersion relations of zigzag ribbons with $N_a=8$, $10$, and $16$. 
  Shaded areas represent the phonon dispersion relations of a two-dimensional graphitic 
  sheet projected onto the zigzag axis. Edge-localized phonon modes are indicated by 
  arrows.}
  \label{fig:5}
  \caption{Atomic displacements of edge-localized phonon states at 
  $k|{\mbox{\boldmath $z$}}|/\pi=1$ in the out-of-plane modes of zigzag ribbon with 
  $N_a=6$ for (a) the transverse acoustic mode and (b) the lowest-lying optical mode. 
  Solid circles represent edge carbon atoms.}
  \label{fig:6}
  \caption{(a) Low-temperature phonon-derived thermal conductance in graphitic ribbon. 
  Open circles denote armchair ribbon ($N_a=9$, $11$, and $15$), and solid circles denote 
  zigzag ribbon ($N_z=8$, $10$, and $16$). (b) Thermal conductances of armchair ($N_a=11$ 
  and $15$) and zigzag ($N_a=10$ and $16$) ribbon as a function of temperature scaled by 
  the energy gap of the lowest optical mode. Broken curve represents a universal curve 
  given by Eq.~(\ref{eq:scale}).}
  \label{fig:7}
\end{figure}

\end{document}